\documentclass[twocolumn,floatfix,superscriptaddress,amsmath,showpacs,showkeys,aps,prb]{revtex4-2}
\usepackage[utf8]{inputenc}
\usepackage[final]{graphicx}
\usepackage{bm}
\usepackage[normalem]{ulem}
\usepackage[usenames]{color}
\usepackage{times}
\usepackage{amsmath}
\usepackage{amsfonts}
\usepackage{mathtools}
\usepackage{verbatim}
\usepackage{amssymb}
\usepackage{xcolor}
\usepackage{hyperref}
\usepackage{physics}
\usepackage{wrapfig}
\usepackage{subfigure}
\DeclareGraphicsExtensions{.eps}

\begin{document}

\title{Critical and non critical non-Hermitian topological phase transitions in one dimensional chains}
\author{Rui Aquino}
\affiliation{Departamento de Física Teórica,
Universidade do Estado do Rio de Janeiro, Rua São Francisco Xavier 524, 20550-013,  
Rio de Janeiro, Brazil}
\author{Nei Lopes}
\affiliation{Departamento de Física Teórica,
Universidade do Estado do Rio de Janeiro, Rua São Francisco Xavier 524, 20550-013,  
Rio de Janeiro, Brazil}
\author{Daniel G. Barci}
\affiliation{Departamento de Física Teórica,
Universidade do Estado do Rio de Janeiro, Rua São Francisco Xavier 524, 20550-013,  
Rio de Janeiro, Brazil}
\date{\today}

\begin{abstract}
In this work we investigate non-Hermitian topological phase transitions using real-space edge states as a paradigmatic tool. We focus on the simplest non-Hermitian variant of the Su-Schrieffer-Hegger  model, including a parameter that denotes the degree of non-hermiticity of the system. We study the behavior of the zero energy edge states  at the non-trivial topological phases with integer and semi-integer topological winding number, according to the distance to the critical point. We obtain that depending on the parameters of the model the edge states may penetrate into the bulk, as expected in Hermitian topological phase transitions. We also show that, using the topological characterization of the exceptional points, we can describe the intricate chiral behavior of the edge states across the whole phase diagram. Moreover, we characterize the criticality of the model by determining the correlation length critical exponent,  directly from numerical calculations of the penetration length of the zero modes edge states.
\end{abstract}

\maketitle

\section{Introduction}
\label{Sec:Introduction}

Non-Hermitian (NH) Hamiltonians are widely used in effective descriptions of a variety of phenomena. Photonic systems~\cite{Lee2009,Xu2016,Doppler2016,Hahn2016,Chen2017}, semimetals, insulators~\cite{Xu2017,Kawabata2019-1,Zyuzin2018,Zhang2021}, electrical circuits~\cite{Yoshida2020,Helbig2020} and interacting systems~\cite{Michishita2020,Fu2017,Pereira2018,Yoshida2018,Yoshida2019,RuiBa2020,Luitz2022} are some examples in which NH phenomena emerges. One prominent feature of NH Hamiltonians is the appearance of  {\em exceptional points} (EPs) in its spectrum~\cite{Rotter_2009, Heiss_2012, Berry_2004}, which arise for specific values of the parameter space. Exceptional points are spectral degeneracies where the eigenvectors coalesce and the NH Hamiltonian becomes defective, inducing remarkable topological properties~\cite{Vyas-2021,Nehra-2022} that have been observed and explored in different experimental set-ups~\cite{Lee2009,Xu2016,Doppler2016,Hahn2016,Chen2017,Zhang2021}. 

NH models exhibit a complex band structure~\cite{Shen_2018}, which gives rise to different physical properties when compared to standard Hermitian systems. One example where these differences become evident is the bulk-boundary correspondence. In the Hermitian case, it states that the non-zero winding number ($W \neq 0$) defined in the bulk determines the existence of gapless zero energy edge states (ZEES), located  at the system's edges (or boundaries). For instance, in the standard Hermitian one-dimensional Su-Schrieffer-Hegger (SSH) model~\cite{SSH_1979} if the winding number is trivial, i.e., $W = 0$, there are no ZEES and therefore, the system is namely a usual insulator. On the other hand, 
for specific values of the parameters, the system undergoes a topological phase transition. At the latter, the system is  known as a topological insulator, where $W = 1$. As a consequence,  there will be a pair of ZEES at the edges of the system (one at the left side and one at the right side of the chain)~\cite{asboth2016short}. In other words, the bulk of the system remains an insulator, while its edges can carry an electric current in the non-trivial topological phase, which makes it so attractive from the technological point of view. By contrast, for NH systems, it is well-know that the winding number could be fractionary~\cite{Chen2018}, along with the fact that the bulk-boundary correspondence fails~\cite{Helbig2020}. A detailed  review,  with some attempts to reestablish this correspondence, can be found in Ref.~\onlinecite{Bergholtz-2021}.

Although a topological phase transition is essentially discontinuous, in the sense that the winding number jumps discontinuously at the transition,  the penetration length of the edge states continuously diverges as a power-law with a well defined exponent. In this sense, at least in the Hermitian case, this kind of  transition has {\em critical} properties\cite{Nei2019,Mucio_book_chapther2020,Nei_2021}.
Moreover, the interest in the critical properties  of NH systems is continuosly growing. For instance, analysis on the thermodynamical properties~\cite{Cristiane2020} as well as  computations of the fidelity susceptibility~\cite{Sun2021} have been made. Disorder effects have been also investigated~\cite{Bao_2021}. In particular, different works have indicated {\em directly or indirectly} the importance of the EPs on the criticality of these NH systems~\cite{Kunst2018,Comaron2020,Rahul2022}.

In order to contribute to this issue, in this work we analyze an extended SSH model, considering  asymmetric hopping~\cite{Cristiane2020,Chen2018}, in order to investigate how the ZEES behave near topological phase transitions and how they  are related to the EPs and their chirality. 

By means of numerical calculations, we compute the topological phase diagram, the complex gap structure and the edge states of the model. Our results show that, depending on the parameters, near the topological transitions, the ZEES may penetrate into the bulk, as expected from Hermitian topological phase transitions~\cite{Nei2019,Mucio_book_chapther2020,Nei_2021}. Surprisingly, the ZEES are not sensitive to the transition from $W = 1/2$ to $W = 1$ winding numbers, that is proper of NH systems. This fact indicates a non-critical character of this particular NH transition. In addition, from the spectrum analysis, we show that one can also characterize the NH topological phase transitions by means of the structure of the complex energy gaps.

Interestingly, from the topological phase diagram, we find that at the non-trivial topological state with $W =1$~(dark rhombus in Fig.~\ref{fig:phase-diagram}) the system exhibits two different  behaviors for the ZEES, depending on the parameter space of the model. One of them is essentially equivalent to the usual Hermitian  behavior, while the other is proper of NH systems and is known as NH skin effect~\cite{Yao-2018,Thomale-2019}, where the zero energy modes accumulate in the boundary. Through a careful analysis of the topological characterization of the EPs, which emerge in the spectrum, we describe each one of these edge state behaviors and their chirality.

Moreover,  we characterize the criticality of the system through the identification of the correlation length critical exponent ($\nu$) of the topological transition from the numerical calculation of the penetration length of the ZEES as a function of the distance to the topological transition point, using  concepts of scaling theory and critical phenomena~\cite{Griffith2018,Sun2017,Kempkes2016,Quelle2016,Nei2019,Nei_2021}.

The paper is organized as follows: In Sec.~\ref{Sec:Non-Hermit-SSH} we present the NH SSH model including a real parameter that encodes the degree of non-hermiticity of the system. In Sec.~\ref{Sec:Topology} we describe the topological aspects of the model and characterize the EPs that emerge in NH systems. In Sec.~\ref{Sec:PD} we obtain the topological phase diagram as well as investigate the behaviors of the EPs, ZEES, and chirality depending on the parameters of the model. In Sec.~\ref{Sec:Topological_surface_states} we calculate the critical exponents that characterize the NH topological transitions and consequently, we identify the universality class of the model. Finally, in Sec.~\ref{Sec:discussions} we conclude and make some remarks about our main results.

\section{The SSH model with asymmetric hopping}
\label{Sec:Non-Hermit-SSH}

\begin{figure}[b]
	\begin{center}
		\includegraphics[width=\columnwidth]{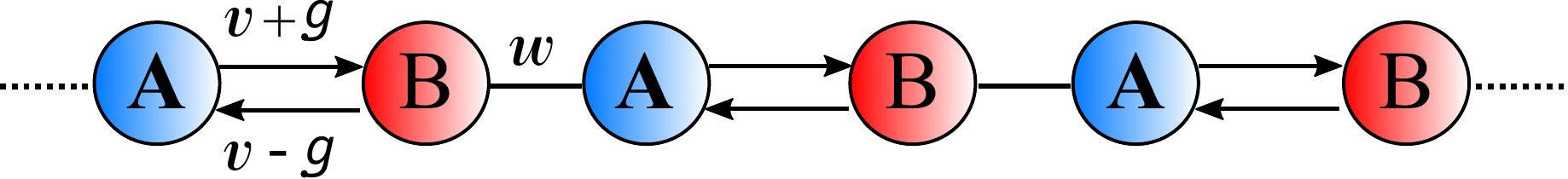}
	\end{center}
	\caption{One-dimensional chain of the non-Hermitian SSH model with asymmetric intracell hopping. Each unitary cell contains a pair of sublattices A and B, the blue and red spheres, respectively. $v$ is the hopping term within the unitary cell, while $w$ is the hopping term out of the unitary cell. The real parameter $g$ measures the degree of non-hermiticity of the model. When $g=0$ we recover the standard Hermitian SSH model.}
	\label{fig:SSH}
\end{figure}

We will analyze the simplest NH variant of the one-dimensional SSH model. In this modified version of the SSH model, we introduce a real parameter ($g$), which denotes an asymmetry in the intracell hopping ($v$). As usual, there is also the term that encodes the intercell hopping ($w$). In Fig.~\ref{fig:SSH} we show the schematic representation for the one-dimensional chain of the non-Hermitian SSH model. 

The Hamiltonian reads~\cite{Cristiane2020,Chen2018},
\begin{align}
H = \sum_n \big[ (v - g) a^\dagger_n b_n &+ (v+g) b^\dagger_n a_n + \nonumber \\
& + w ( a^\dagger_{n+1} b_n + b^\dagger_n a_{n+1}) \big]~,
\label{eq.: hamiltonian}
\end{align}
where $a^\dagger_n$, $(a_n)$ and $b^\dagger_n$, $(b_n)$ are the creation (annihilation) operators at the $n$-th $A$ and $B$ site, respectively. Note that the parameter $g$ acts as the degree of non-hermiticity of the system and when $g = 0$ we recover the SSH model~\cite{SSH_1979}.

Performing a Fourier transformation, and using $\psi = (b_k,a_k)^T$, one can rewrite the Hamiltonian in Eq.~(\ref{eq.: hamiltonian}) as follows, 
\begin{align}
H = \sum_k \psi^\dagger_k h(k) \psi_k
\end{align}
with the one-particle Hamiltonian in the form,
\begin{align}
h(k) = h_x(k) \sigma_x + h_y(k) \sigma_y, \label{eq:one-particle-h}
\end{align}
where $\sigma_{x,y}$ are the Pauli matrices and the components $h_x$ and $h_y$ are given by~\cite{Cristiane2020,Chen2018},
\begin{align}
h_x(k) &= v + w \cos(k) \label{eq: dx} \\
h_y(k) &= w \sin(k) - i g.\label{eq: dy}
\end{align}
For simplicity, we are measuring distances in terms of the lattice spacing $(a)$, or we simply set $a=1.0$.
Note, again, that if $g=0$ we recover the Hermitian SSH model~\cite{SSH_1979}, as expected. Due to the fact that $v, w \in \mathbb{R}$, explicitly putting Eq.~(\ref{eq: dx}) and Eq.~(\ref{eq: dy}) into Eq.~(\ref{eq:one-particle-h}), one can see how the parameter $g$ breaks the hermiticity of the Hamiltonian. Computing the dispersion relation of Eq.~(\ref{eq:one-particle-h}) we find $E_\pm = \pm \sqrt{h_x^2 + h_y^2}$, as usual. 

When dealing with Hermitian systems, hermiticity ensures real eigenvalues as well as orthogonal eigenvectors. On the other hand, for NH systems the eigenvalues may be complex and the Hamiltonian allows a complete biorthonormal system of eigenvectors when it is diagonalizable~\cite{Mostafazadeh2002}. In conclusion, the Hilbert space of the system support a biorthogonal basis $\braket{u_\pm}{{\tilde u}_\pm} = \delta_{ij}$, where $i,j = +,-$ and ${\tilde u}$ is the dual of $u$. This basis are defined by,
\begin{align}
h(k)\ket{u_\pm} = E(k)\ket{u_\pm} \\
h^\dagger(k) \ket{{\tilde u}_\pm} = E^*(k) \ket{{\tilde u}_\pm}
\end{align}
where the eigenvectors are given by
\begin{align}
\ket{u_\pm} = &\left( \begin{array}{c}
\pm\, e^{-i \phi(k)} \\
1
\end{array}\right), ~~ 
\ket{{\tilde u}_\pm} = \left( \begin{array}{c}
\pm\, e^{-i \phi^*(k)} \\
1
\end{array} \right)
\label{eq: eigenvec}
\end{align}
and
\begin{align}
\phi(k) = \tan^{-1}\,\left( \frac{h_y}{h_x}\right) \label{eq: phi}~.
\end{align}

Note that $h_y(k)$ is a complex function that depends on $k$, see Eq.~(\ref{eq: dy}). So, the angle $\phi(k)$ in Eq.~(\ref{eq: phi}) is, in general, complex. 

It is worth to emphasize that due to the complex character of the dispersion relation, $E(k)$, the spectrum
has a more involved structure than in the Hermitian case.  So, when dealing with NH systems, we have two independent definitions of complex energy gaps, which are given by the following statements~\cite{Kawabata2019-1},
\begin{itemize}
\item Point gap - A NH system presents a point gap if and only if it is invertible and has nonzero eigenenergies, i.e., $\det H(k) \neq 0$ and $E(k) \neq 0~ \forall~ k$, respectively. 
\item Real (Imaginary) Line gap - A NH system presents a line gap in the real (imaginary) part of its spectrum if and only if it is invertible and the real (imaginary) part of the eigenenergies are nonzero, i.e., $\det H(k) \neq 0$ and $\Re E(k) \neq 0 ~(\Im E(k) \neq 0)~ \forall ~k$, respectively.
\end{itemize}

\begin{figure}[t]
	\begin{center}
		\includegraphics[width=\columnwidth]{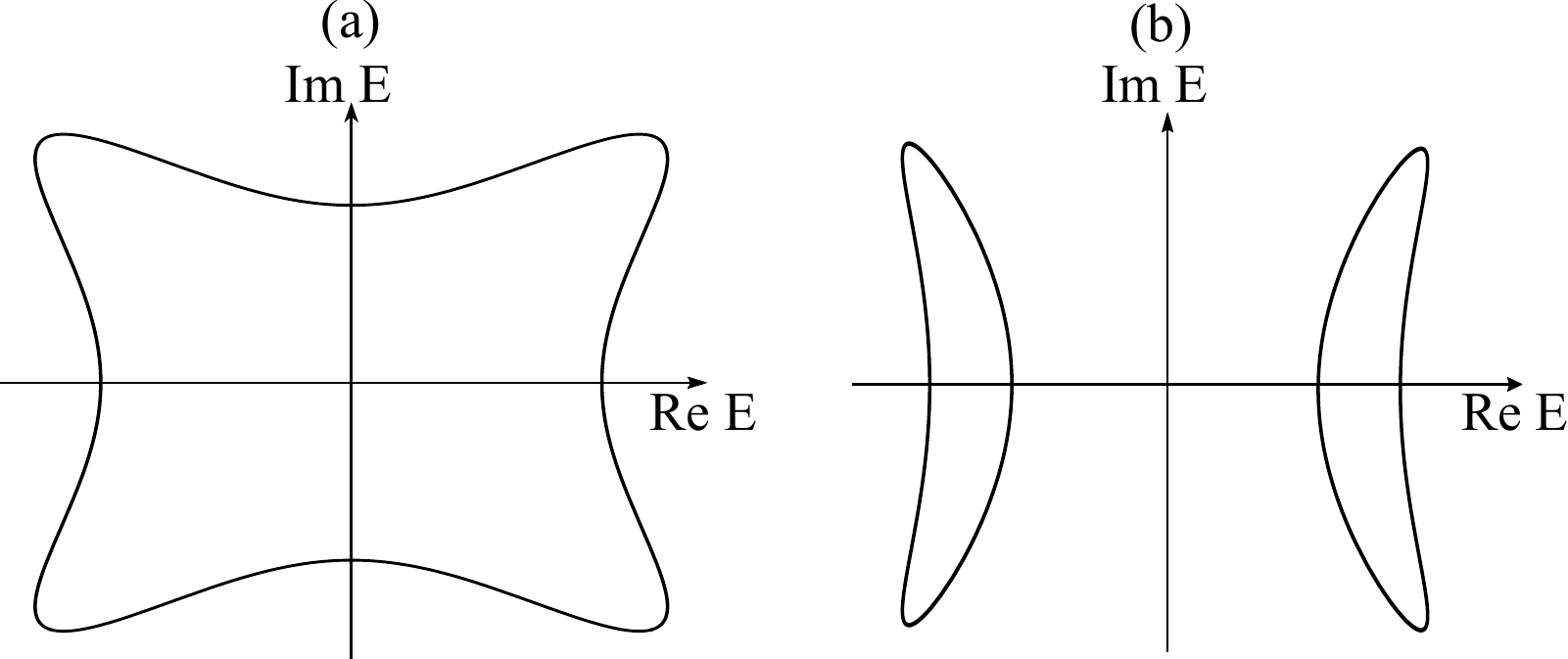}
	\end{center}
	\caption{(a)~we show the complex energy gap for $v = 0.3$ and $g = 1.0$. For these parameters we have a point gap spectrum.~(b) we show the complex real line gap for $v = -0.1$, $g = 0.6$. We fix $w =1.0$ in both panels.}
	\label{fig:Gaps}
\end{figure}

The structure of the complex energy gaps are extremely important to the topological characterization of NH systems, as we will discuss latter. In Fig.~\ref{fig:Gaps} we plot some example of gaps that appear in the NH SSH model with asymmetrical hopping, and in Sec.~\ref{Sec:PD} we will give more details about the complex energy gaps through the topological phase diagram of the model.

\section{Topology and Exceptional Points}
\label{Sec:Topology}

In the usual ten-fold classification of Hermitian topological phases~\cite{Schnyder2008,Altland1997}, three symmetries are required to its description, which are time-reversal, particle-hole and chiral symmetry. On the other hand, in the NH classification~\cite{Kawabata2019-2}, symmetries ramifies due to the distinction between transposition and complex conjugation in NH Hamiltonians. In fact, chiral symmetry, which gives rise to topological properties for the Hermitian SSH model, is distinct from the sublattice symmetry, although they are equivalent in Hermitian physics. The NH SSH model, defined in last section, belongs to Class A with an additional Sub-lattice Symmetry (SLS), as defined in~\cite{Kawabata2019-1},
\begin{align}
{\cal S} h(k) {\cal S}^{-1} = - h(k), ~~~~ {\cal S}^2 = 1~,
\end{align}
where ${\cal S}$ in general is an unitary matrix  of the non-Hermitian 38-fold classification~(Note that the model does not have chiral symmetry, so it does not respect any AZ$^\dagger$ symmetry).

For the Hamiltonian given by Eq.~(\ref{eq:one-particle-h}), ${\cal S} \coloneqq \sigma_z$. Additionally, we have the winding number defined as~\cite{Chen2018},
\begin{align}
W = \frac{1}{2 \pi} \oint \partial_k \phi \, dk \label{eq:w}
\end{align}
where $\phi$ was previously defined in Eq.~(\ref{eq: phi}) and the integral is taken along a loop with $k$ from $0$ to $2\pi$. 

It is interesting to point out that if we compute the winding number for the Hamiltonian in Eq.~(\ref{eq:one-particle-h}), we found that $W$ is quantized as $\mathbb{Z}/2$, rather than $\mathbb{Z}$, as we have in the Hermitian SSH model. This fractional winding number, although unexpected, is in agreement with others topological characterizations of NH systems. In Ref.~\onlinecite{Kawabata2019-1}, the class A with SLS defines a $\mathbb{Z}$ winding number for systems with line gap spectrum, and a $\mathbb{Z}\oplus\mathbb{Z}$ winding number for point-gap spectrum (see Table VI of the reference~\onlinecite{Kawabata2019-1}). As pointed out in Ref.~\onlinecite{Chen2018}, we can rearrange Eq.~\ref{eq:w} as $2W = W_1 + W_2$, where $W_1$ and $W_2$ are $\mathbb{Z}$ topological invariants, so we can recover all the results of the topological characterization of Ref.~\onlinecite{Kawabata2019-1} using $W_1$ and $W_2$ as defined in Ref.~\onlinecite{Chen2018}. Interestingly, this $\mathbb{Z}/2$ winding number was recently realized in quantum simulators~\cite{Zhang2021}. 

Due to the fact that Eq.~(\ref{eq:w}) just have contributions from the real part of the angle $\phi(k)$, we can geometrically interpret the $\mathbb{Z}/2$ quantization as the manifestation of the encircling of the EPs of the model in the $\{\Re[h_x],\Re[h_y]\}$ space. To understand it, let us find the location of the EPs by computing the zeros of the dispersion relation, i.e., $h_x^2 + h_y^2 = 0$, or equivalently,
\begin{align}
&\Re[h_x] = - \Im[h_y] ~~ \text{and} ~~ \Re[h_y] = \Im[h_x] \\
&\Re[h_x] = \Im[h_y] ~~ \text{and} ~~ \Re[h_y] = - \Im[h_x]~.
\end{align}

\begin{figure}[t]
	\begin{center}
	\includegraphics[width=\columnwidth]{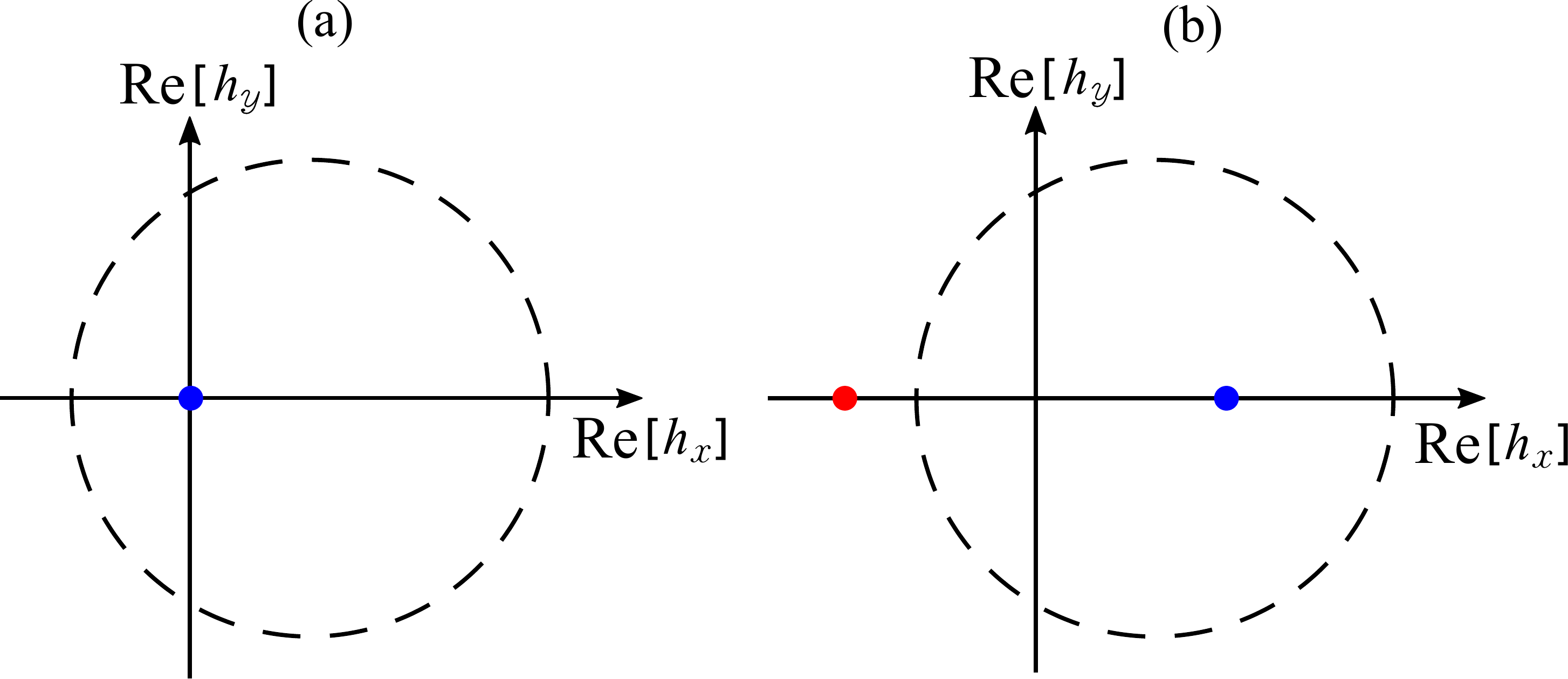}
	\end{center}
	\caption{$\{\Re[h_x],\Re[h_y]\}$ plane and the encircling (dashed) of the exceptional points (red and blue dots) fixing $w=1.0$. (a)~We take $g = 0$ and $v = 0.5$, i.e., the non-trival topological phase of the standard Hermitian SSH model with $W = 1$. (b)~Non-Hermitian topological phase with $W = 1/2$, where $g = 0.8$ and $v = 0.5$. Note that in the Hermitian case the exceptional points are both located at the origin, while in the non-Hermitian case the exceptional points are shifted along the $\Re[h_x]$ axis and only one is encircled.}
	\label{fig:windings}
\end{figure}

So, we can conclude that each one of the two EPs appear in the ordered pairs $(-\Im[h_y],\Im[h_x])$ and $(\Im[h_y],-\Im[h_x])$. In Fig.~\ref{fig:windings} we show some examples of the different phases of the model described by the EPs. In this figure, the dashed circle are the geometrical points of ($\Re[h_y(k)], \Re[h_x(k)]$),  when $k$ goes form $0$ to $2\pi$ for a fixed values of the parameters. 

In Fig.~\ref{fig:windings}~(a) we have both EPs merging at the origin, that is, $(h_x,h_y) = (0,0)$, which can be recognized as the non-trivial topological phase of the usual Hermitian SSH model with $W=1$, while in Fig.~\ref{fig:windings}~(b) we have an exclusive NH topological phase where $W=1/2$. Note that for the latter, only one EP is encircled.

In general, we can conclude that when two EPs are encircled, $W=1$. On the other hand, when only one of them is winded, $W=1/2$. Finally, when no EPs is encircled, $W=0$. We will use the trajectories of the EPs to describe the chirality and the phase transitions of the model.

\section{Topological phase diagram, chirality and edge states}
\label{Sec:PD}
\begin{figure}[b]
	\begin{center}
		\subfigure[]
		{\label{fig:phase-diagram}
			\includegraphics[width=0.47\textwidth]{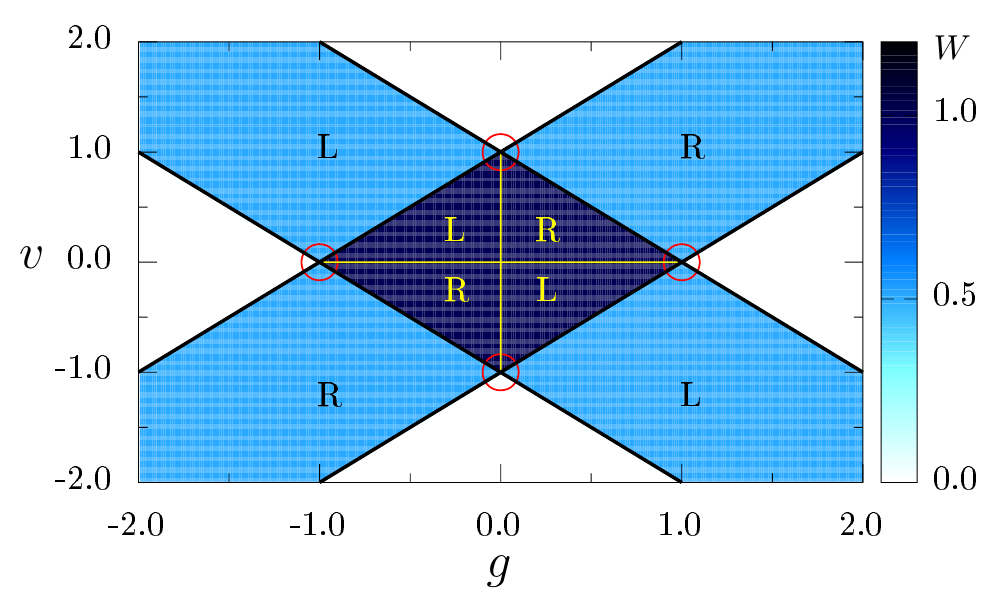}}
		\subfigure[]
		{\label{fig:complex-gaps}
			\includegraphics[width=0.47\textwidth]{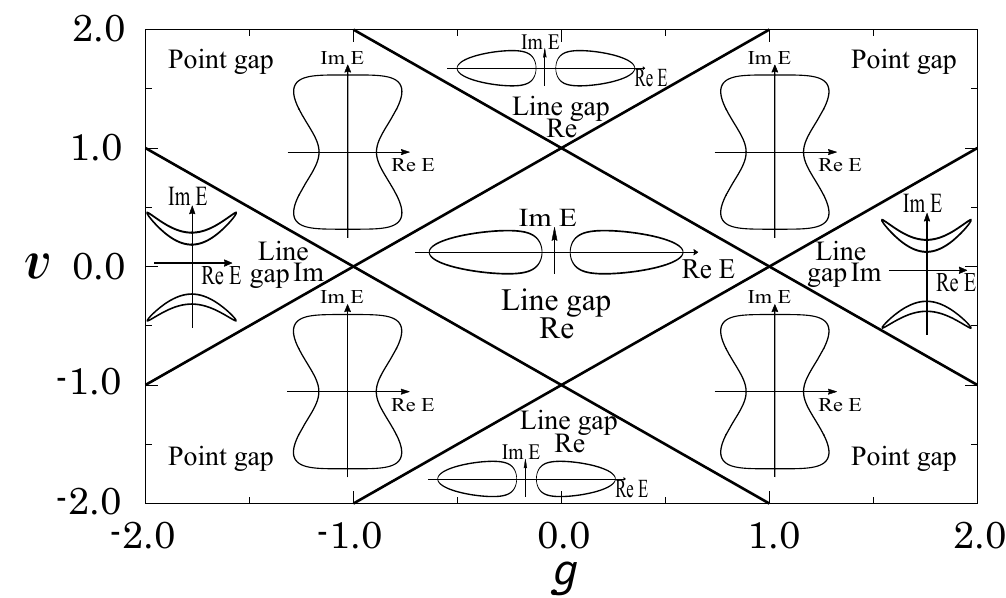}}
	\end{center}
	\caption{(a) Phase diagram of the non-Hermitian SSH model obtained from the calculation of Eq.~(\ref{eq:w}) fixing $w=1.0$. We can see three topological phases characterized by $W = 0, 1/2$ and $1$, see colorbar. The chirality of the model is represented by the presence of left (L) or right (R) zero energy edge states. Within the vertical and horizontal continuous lines (yellow) there are edge states in both sides (left and right) of the chain as well as it penetrate into the bulk of the system as we approach the transition point. Note that these continuous lines divide the phase diagram in four quadrants inside the rhombus with $W = 1$, where the chirality depends on the parameters of the model even at the non-trivial topological state with an integer topological winding number. (b) Complex gaps structure in each phase of the system.} 
	\label{fig:phase_diagram_non_Herm_SSH}
\end{figure}

The topological phase diagram of the NH SSH model can be obtained through the calculation of the topological invariant winding number given by Eq.~(\ref{eq:w}). The phases that emerge on the model are characterized by $W = 0, 1/2$ and $1$, as shown in Fig.~\ref{fig:phase-diagram}. The central dark rhombus is characterized by a winding number $W=1$, the blue zones out of the rhombus are characterized by $W=1/2$ while the white zones are in the trivial topological phase  $W=0$. On the other hand, in Fig.~\ref{fig:complex-gaps} we depict the complex gaps structure of the model. Note that the central rhombus exhibits a real line gap, while the phases with fractional winding number present a point gap. Interestingly, the $W = 0$ regions have two different line gaps. The upper and lower trivial regions ($W = 0$) have a real line gap, while the left and right $W = 0$ regions have imaginary line gaps.

Moreover, also note from Fig.~\ref{fig:complex-gaps} that within the line $g=0$, i.e., in the Hermitian case, the topological phase transitions only exhibit the real line gap, as expected. However, when dealing with NH systems, i.e., for $g \neq 0$, one can see that  when the NH system undergoes a topological phase transition the gap structure necessarily changes. Thus, we can also characterize the NH system through its complex gap structure.

In the following we will discuss the chiral aspects of the topological phase diagrams presented in Fig.~\ref{fig:phase-diagram} and the character of the ZEES as we approach the NH topological phase transitions.

\subsection{Chirality}
\label{Sec:chiral}

In the topologically trivial phase with $W = 0$ the system does not have ZEES and behaves as a conventional insulator, as expected. At the exclusive NH topological phase, where $W=1/2$, the two ZEES are accumulated at the edge (left (L) or right (R) side) of the chain, which is a direct manifestation of the NH skin-effect~\cite{Yao-2018,Bergholtz-2021}. Moreover, the accumulation of the ZEES respect the chirality depicted in Fig.~\ref{fig:phase-diagram}. This behavior have a direct relation with the EPs and its position in the $\{\Re[h_x],\Re[h_y]\}$ space. 

\begin{figure}[t]
  \includegraphics[width=\columnwidth]{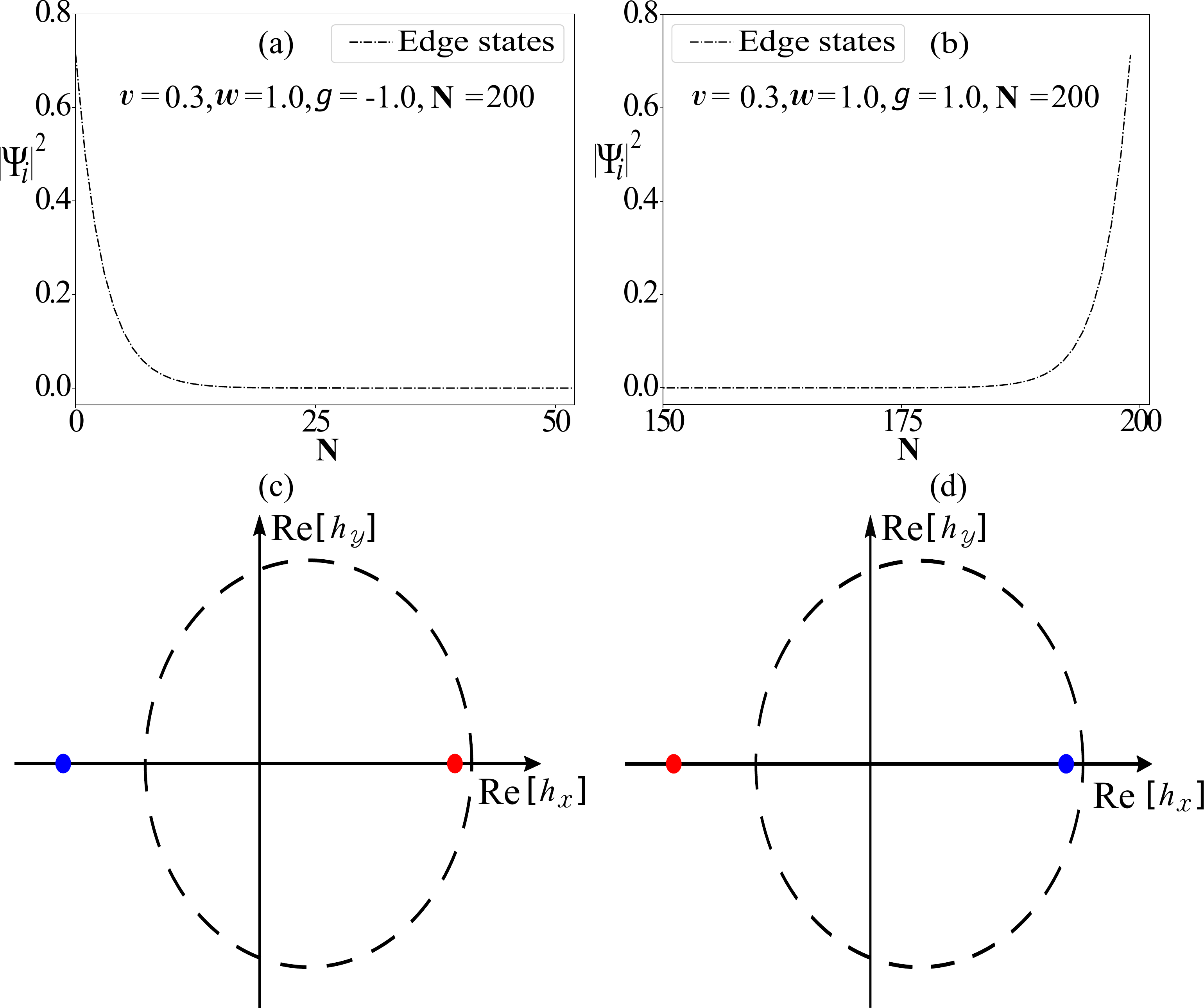}
  \caption{Chirality of the zero energy edge states as we enter in the fractional topological phase with $W=1/2$ from the trivial one. (a)~Left edge states at the semi-integer topological phase with left chirality. (b)~Right edge states at the semi-integer topological phase with right chirality. (c)~The left chirality can be determined by the EP (red dot) that enters first in the unit circle (dashed), which is in the $\{\Im[h_y],-\Im[h_x]\}$ position. (d)~The right chirality can be determined by the EP (blue dot) that enters first in the unit circle (dashed), which is in the $\{-\Im[h_y],\Im[h_x]\}$ position. In~(c) and~(d) we fix $v = 0.3$ and take $g = -1.2$ and $g = 1.2$, respectively. We show a restrict region of the chain in (a) and~(b) since there is only left or right edge states.} 
	\label{fig:fractional-phase-chirality}
\end{figure}

The chirality of the ZEES are determined by which EP enters the unit circle first. For instance, in the topological phase diagram of the Fig.~\ref{fig:phase-diagram} the fractional phases with left (L) or right (R) chirality exhibit left or right ZEES, respectively, see Figs.~\ref{fig:fractional-phase-chirality}~(a) and~(b). In Fig.~\ref{fig:fractional-phase-chirality}~(c) we show that the EP in position $\{\Im[h_y],-\Im[h_x]\}$ (red dot in all the figures) enters first in the unit circle (dashed) as we approach the semi-integer topological phase with left chirality from the trivial one, while Fig.~\ref{fig:fractional-phase-chirality}~(d) shows that the EP in position $\{-\Im[h_y],\Im[h_x]\}$ (blue dot in all the figures) enters first in the unit circle (dashed) as we approach the semi-integer topological phase with right chirality from the trivial one.

Surprisingly, the topological phase transition from $W = 1/2$ to $W = 1$, and vice versa, does not affect the chirality of the edge states, as can be seen in Fig.~\ref{fig:phase-diagram}. This indicate an odd character of this transition, and it will be explored in the discussion of the ZEES behavior in next section. Moreover, inside the rhombus with $W=1$, we find an interesting behavior for the chirality of the ZEES. When we cross the $g=0$ line, the chiralites of the ZEES change. It can be understood by the shift of the EPs around the origin of the $\{\Re[h_x],\Re[h_y]\}$ space. Within this line, we have the Hermitian SSH model~\cite{SSH_1979}, so this crossover passing through the Hermitian limit of the model affects the chirality. Analogously, if we cross the $v = 0$ line we also have a change in the chirality. We can track this by the symmetry that the model has in the ``parity'' transformation of the parameters $(v,g) \to (-v,-g)$. In fact, the EPs positions are the same if we perform this change in the parameters.

The chiral behavior of the ZEES in the fractional phase has been previously reported in Ref.~\cite{Chen2018}. However, to the best of our knowledge, this two lines for $g=0.0$ and $v=0.0$ inside the rhombus with $W = 1$ and the relation of the chirality in this region with the EPs were not clarified until now.

\subsection{Edge states and exceptional points}
\label{Sec:EdgeStates}

\begin{figure}[t]
  \includegraphics[width=\columnwidth]{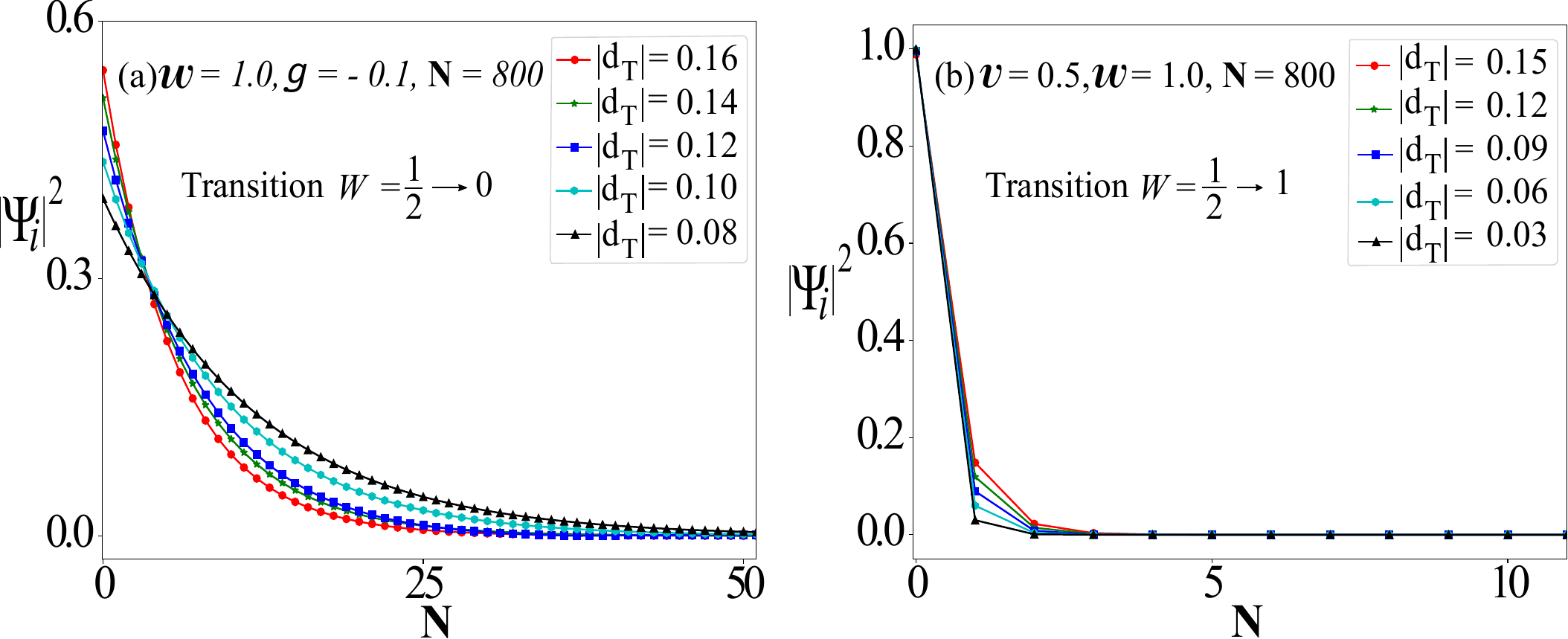}
  \caption{Penetration of the zero energy edge states as we approach the topological transition point varying $v$ and $g$, respectively. (a)~The edge states penetrate into the bulk from the semi-integer to trivial topological transition. (b)~The edge states get more localized for the semi-integer to integer transition as we approach the criticality increasing the degree of non-hermiticity of the system. Again, we show only a part of the chain in $x$-axis since we just have edge states at the left side. $d_T$ is the distance to the transition point.} 
	\label{fig:semi-integer-trivial}
\end{figure}

When we approach the transition with $W=1/2 \to W = 0$, i.e., the semi-integer to the trivial topological phase transition, the ZEES penetrate into the bulk, in the same sense of Hermitians topological phase transitions~\cite{Nei2019,Nei_2021}, see Fig.~\ref{fig:semi-integer-trivial}~(a). By contrast, when we approach the topological phase transition with $W=1/2 \to W=1$ we obtain distinct behaviors depending as we approach the transition point. For instance, the ZEES get more localized at the left (or right) side of the system as we approach the transition point increasing the degree of non-hermiticity ($g$), see Fig.~\ref{fig:semi-integer-trivial}~(b).  On the other hand, as we approach the transition decreasing the value of $g$ the ZEES tends to penetrate into the bulk (not shown). In this sense, we confirm that the degree of non-hermiticity of the system is associated with the chirality and acts to localize the ZEES at the edges of the system. Furthermore, the ZEES are not affected if we approach the semi-integer to integer topological transition, and vice versa, considering the same angular coefficient of the chiral regions in the topological phase diagram of Fig.~\ref{fig:phase-diagram} (not shown), which, again, suggests an unusual aspect for the NH topological transitions. In the next section, we will relate this non unique character of this phase transition to a non critical behavior. However, note that we can indeed identify all the topological phase transitions of the model through the changing on the complex spectrum structure, as pointed out before, see Fig.~\ref{fig:complex-gaps}.
\begin{figure}[b]
 \includegraphics[width=\columnwidth]{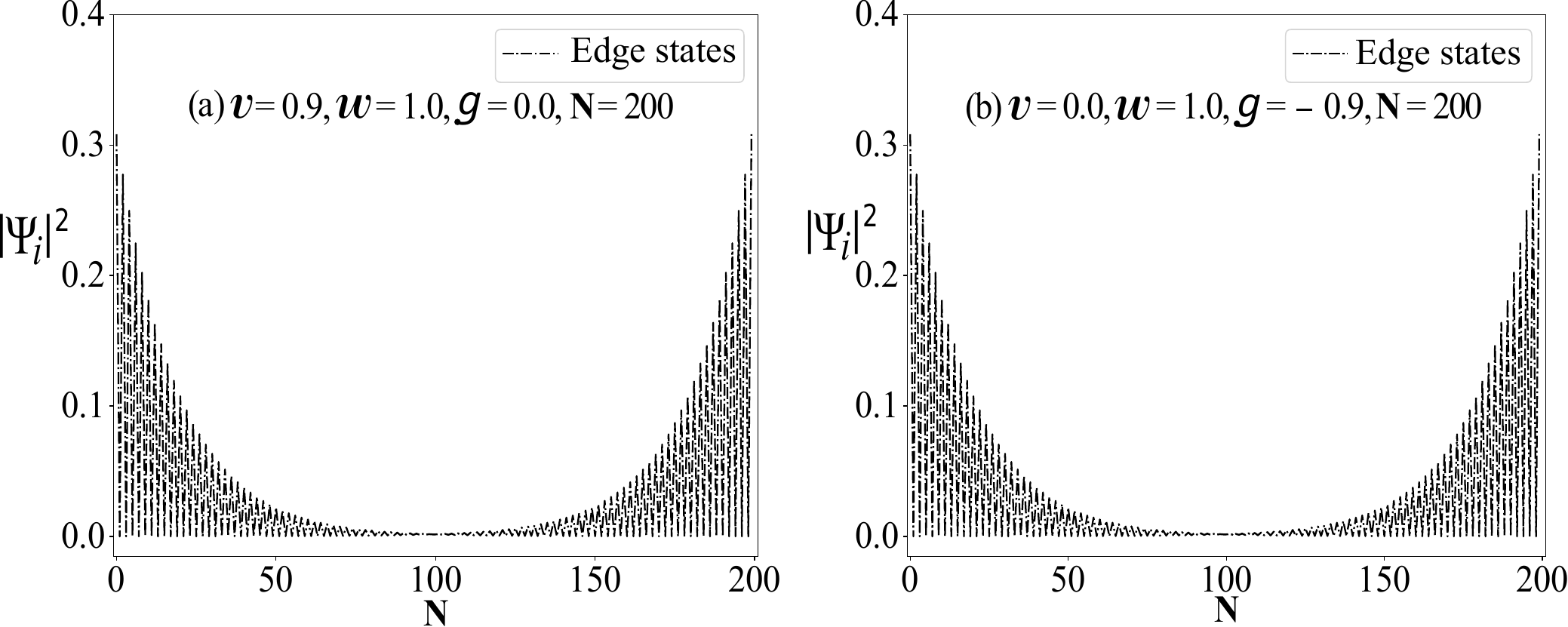}
  \caption{Exponentially decay of the amplitude of the wave function of the zero energy edge modes into the bulk according to the phase diagrams of Fig.~\ref{fig:phase_diagram_non_Herm_SSH}. (a) and (b) Edge states decay for vertical and horizontal continuous lines taking $g=0.0$ and $v = 0.0$ close to the transition point, respectively.}
	\label{fig:edge_states_cross}
\end{figure}

Let us now focus on the rhombus with $W=1$. Note that the continuous vertical line for $g=0$ in Fig.~\ref{fig:phase-diagram} recovers the usual topological transitions presented on the Hermitian SSH model~\cite{SSH_1979, Nei2019}. The ZEES within this line, near the transition, at the upper edge of the rhombus, is shown in Fig.~\ref{fig:edge_states_cross}~(a). The edge states as well as the transition for the bottom edge of the rhombus are completely symmetric, i.e., changing the sign of $v \to -v$, as suggested by the topological phase diagram. 

Surprisingly, within the horizontal continuous line in Fig.~\ref{fig:phase-diagram}, i.e., for $v = 0$, at the non-trivial topological phase with $W=1$ we also obtain ZEES at the both edges of the chain, as shown in Fig.~\ref{fig:edge_states_cross}~(b). This kind of behavior for the exclusive anti-hermitian case can be understood from the Hamiltonian in Eq.~(\ref{eq.: hamiltonian}). Note that when $v=0$ the hopping terms present the same magnitude in modulus, i.e., $|g|$, which implies that the net particle flux is equal for both sides (left and right) of the chain. These two ZEES are completely symmetric and its edge states penetrate into the bulk as we approach to the topological transition. 

\begin{figure}[b]
  \includegraphics[width=\columnwidth]{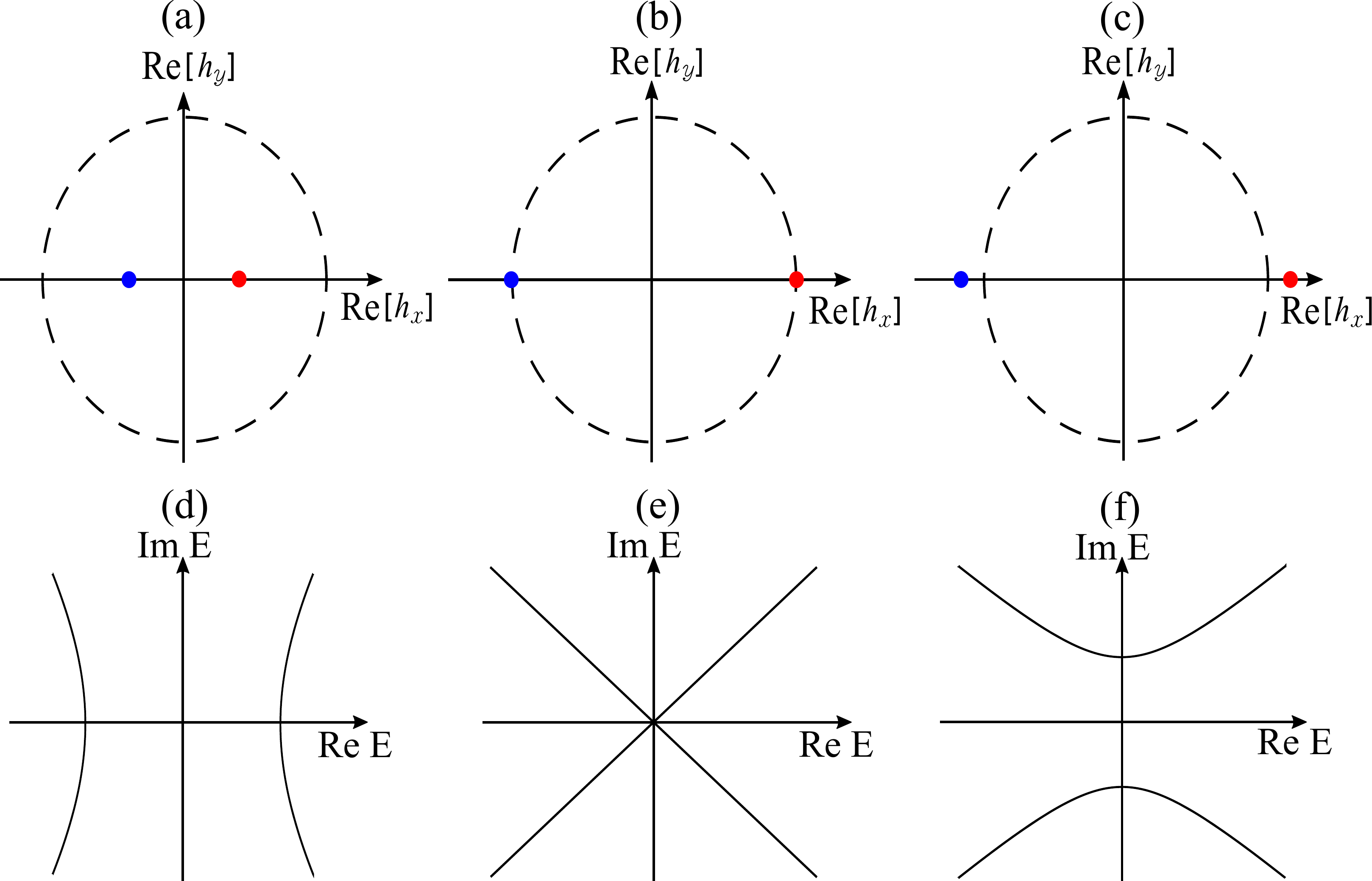}
  \caption{(a),~(b),~(c) The $\{\Re[h_x],\Re[h_y]\}$ parameter space as we approach to the topological transition within the horizontal continuous line with fixed $v = 0.0$ in Fig.~\ref{fig:phase-diagram} for $g = -0.6$, $g = -1.0$ and $g = -1.1$, respectively. The unit circle (dashed) remains fixed at the origin, while the exceptional points move along the $x$-axis. For this case, both exceptional points enter ($W = 1$) or leave ($W = 0$) the unit circle together at the topological transition point. (d),~(e), and (f) The line gaps for the same values of $g$ in~(a),~(b) and (c), respectively, showing the non triviality of this topological phase transitions within this special line.}
	\label{fig:line-v=0}
\end{figure}

This penetration is expected for $g=0.0$, that is, for the Hermitian case, but it is not so trivial for the horizontal line where $v = 0.0$. In this limit, the unit cell is completely anti-Hermitian, however cleary its edge states behaves exactly as the full Hermitian limit of the NH SSH model. If one investigate the trajectories of the EPs in the $\{\Re[h_x],\Re[h_y]\}$ parameter space, we see that, due to the fact that $v = 0.0$, the unit circle is fixed at the origin, and the trajectories of the EPs are perfectly symmetric with respect to the circle.  For the phases with $v\neq 0.0$ and $g \neq 0.0$, the assymetry of the trajectories induces topological phase transitions where one EP at a time enters in the unit circle. This is not the case for $v = 0.0$, i.e., when we approach the topological phase transitions at $g = -1.0$ and $g = 1.0$, the EPs cross the unit circle together. Another interesting point is that the gaps within the line $v=0.0$ change from imaginary to real line gap (and vice-versa) at the topological phase transitions, see Fig.~\ref{fig:complex-gaps}. In this special limit, the edge modes behave identically as in a Hermitian topological phase transition. We show these trajectories of the EPs and its complex gaps in Fig.~\ref{fig:line-v=0}.

On the other hand, any other case outside the continuous vertical or horizontal line at the region with $W=1$ in Fig.~\ref{fig:phase_diagram_non_Herm_SSH}, we recover the NH skin effect, with the two modes accumulated at the left or right edges. In this sense, one can conclude that this two vertical and horizontal lines divide the rhombus where $W=1$ into four quadrants. Again, these results are completely symmetric for even or odd quadrants.

\section{Topological surface states and critical exponents}
\label{Sec:Topological_surface_states}

When a system undergoes a phase transition, in principle, there is only one diverging length, i.e., the correlation length ($\xi$), that dominates the phase transition in proximity to the quantum critical point (QCP). Our main goal, when applying the penetration depth technique~\cite{Nei2019, Nei_2021}, is to identify the diverging behavior of the penetration depth of the zero energy edge modes into the bulk as we approach the topological transition point.

It is possible to get information about the critical exponents from simple scaling assumptions.  For instance,  very near the  phase transition,the singular term of the ground state energy density scales as $f_{s} \propto |t-t_c|^{\nu(d+z)}$,  where $t$ is a generic control  parameter and $t_c$ is the transition point.  $d$ is the dimensionality of the system,  $z$ is the dynamical critical exponent and the critical exponent $\nu$ characterizes the divergence of the correlation length~\cite{Mucio_book_chapther2020}. Moreover, in the one-dimensional non-Hermitian SSH case, the modulus of the spectrum of excitations, very near the topological phase transition, can be written as,
\begin{equation}
\left| E(k)\right|\sim  \sqrt{|t-t_c|^{2\nu z}+k^{2z}}~.
\label{eq:ModEk}
\end{equation} 
As an example,  let us consider for instance $v=0$ and $g\to 1$  in our model.  In this case,  by direct calculation, 
\begin{equation}
\left|E(k)\right|\sim \sqrt{2} \left\{(1-g)^2+k^2 \right\}^{1/4}~.
\end{equation}
Thus,   $\lim_{g\to 1}\left| E(k)\right|\sim k^{1/2} $ and  $\lim_{k\to 0}\left| E(k)\right|\sim (1-g)^{1/2} $.  Comparing with Eq.~(\ref{eq:ModEk}) we identify  $2z=1$ and $2\nu z =1$.  Therefore, we get $\nu=1$. We have checked all the transition lines of our model, obtaining $\nu=1$ in all the transitions from a non-trivial winding to a trivial one.

\begin{figure}[t]
  \includegraphics[width=\columnwidth]{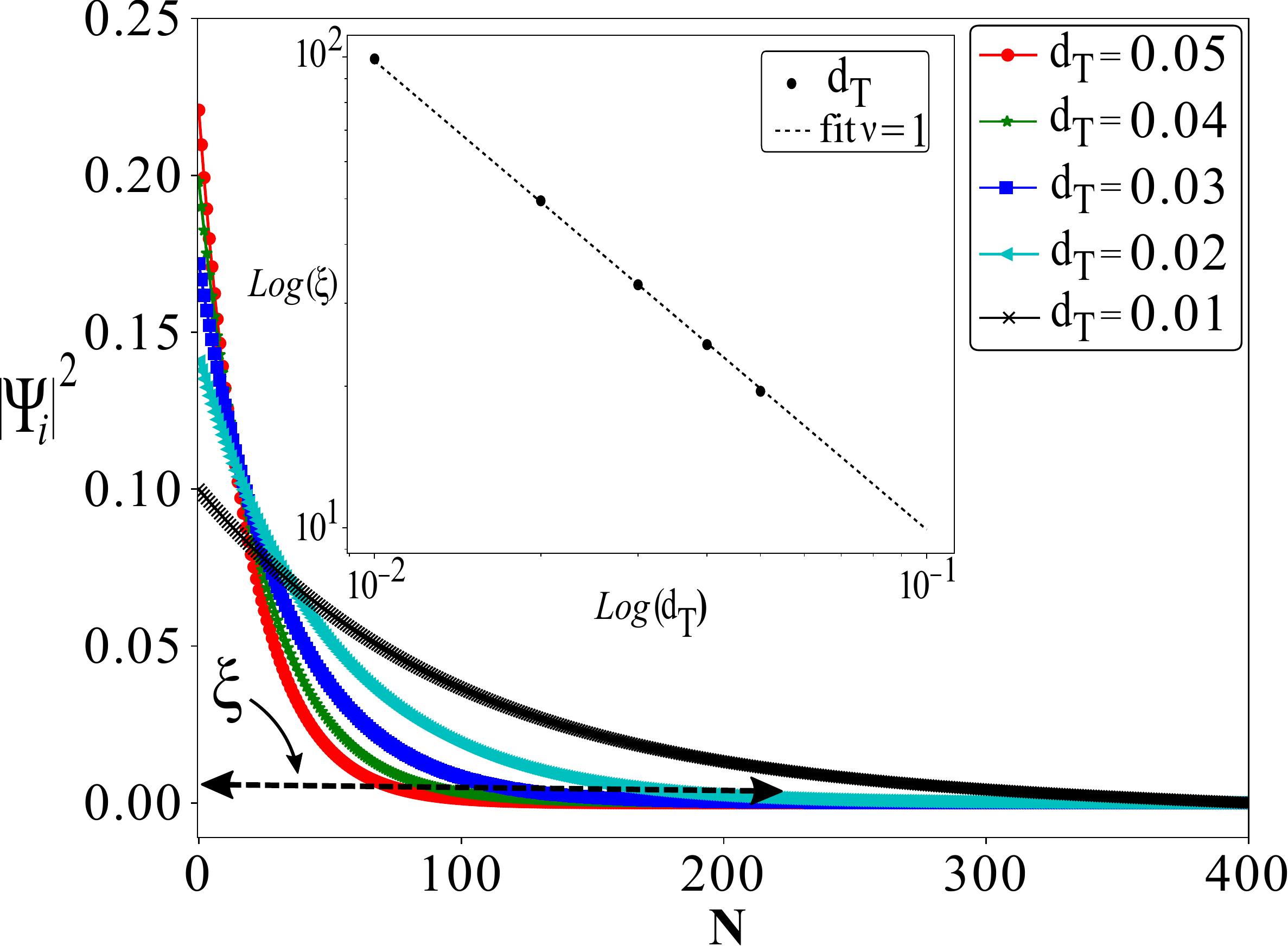}
  \caption{Exponentially decay of the wave function of the zero energy edge modes into the bulk of the system for the one-dimensional SSH model with $N=800$ and fixed $w=1.0$ and $g=0.0$, where $d_{T} = |t-t_c|$, as we approach the transition point. Note that the edge states penetrate into the bulk as we get closer to the critical point ($\xi \rightarrow \infty$). Defining $\xi$ as the correlation length of the topological transition and using the scaling relation $\xi \propto d_T^{-\nu}$ we can identify the correlation length critical exponent $\nu=1$ by means of the angular coefficient of the fitting curve in a Log vs Log plot, see inset.} 
	\label{fig:penetration_depth}
\end{figure}

In this section, we will confirm the scaling hypotesis, by direct numerical calculation  of the edge states in real space.  Since we already know the critical parameters,  given exactly in Fig.  (\ref{fig:phase-diagram}),  instead of  a finite size scaling technique\cite{WangWang-2020},    we  choose an equivalent numerical procedure that has lower computational cost and  have already proved to be very accurate in similar cases~\cite{Nei2019,Mucio_book_chapther2020,Nei_2021}.
In this sense,  diagonalizing numerically the Hamiltonian given by Eq.~(\ref{eq.: hamiltonian}), in real space, one can obtain the eigenvalues and eigenvectors for the one-dimensional chain with $N$ sites. To define the penetration depth as the diverging length of the topological transition, in principle, the ZEES may penetrate into the bulk as we approach the transition point. In addition, we consider the distance to the transition point in the form $d_{T}=|t-t_c|$, where $t$ is a generic control parameter and $t_c$ is the transition point. Since the wave function of the zero energy edge modes decay exponentially as we approach the topological transition point we can identify the correlation length $\xi$ as the diverging length of the topological transition and, therefore, we can use the scaling relation $\xi \propto |t-t_c|^{-\nu}$ to obtain the value of the correlation length critical exponent $\nu$ by means of a Log vs Log plot~\cite{Nei2019, Nei_2021}, as shown in Fig.~\ref{fig:penetration_depth}. 

Following this procedure we can investigate the topological phase diagrams of the non-Hermitian SSH model, given by Fig.~\ref{fig:phase_diagram_non_Herm_SSH}, in order to calculate the values of $\nu$ for the topological transitions, and consequently characterize the criticality of the SSH model with asymmetric hopping. We calculate the value of the critical exponent $\nu$ for the topological transition points at the edges of the rhombus with $W=1$, see red circles in Fig.~\ref{fig:phase-diagram}. We approach the topological transition points from different paths described by the superscript arrows in Table~\ref{tableI} and our numerical results show that these zero energy edge states penetrate into the bulk, which makes the penetration depth method~\cite{Nei2019,Nei_2021} suitable. Note that all values of $\nu$ are very close to the unit, meaning that even for NH topological transitions the universality class of the system is the same of the Hermitian SSH model for the topological transitions at the edges of the rhombus with $W = 1$.

\begin{table}[h]
\begin{tabular}{|c|c|c|}
\hline
    $(v,g)$ &$W$ &$\nu$ \\
\hline \hline
    $(0.95,0.0)$  &$1$	  & $\nu_v^{\uparrow}= 0.998$ \\

    $(-0.95,0.0)$ &$1$   & $\nu_v^{\downarrow}= 0.998$ \\

    $(0.0,0.95)$  &$1$   &$\nu_h^{\rightarrow}= 0.998$ \\

    $(0.0,-0.95)$ &$1$   &$\nu_h^{\leftarrow}= 0.998$  \\

    $(1.0,-0.15)$ &$1/2$ &$\nu_h^{\rightarrow}= 0.979$ \\

    $(0.23,-1.0)$ &$1/2$ &$\nu_v^{\downarrow}= 0.944$ \\

    $(-1.0,0.15)$ &$1/2$ &$\nu_h^{\leftarrow}= 0.979$ \\

    $(-0.23,1.0)$ &$1/2$ &$\nu_v^{\uparrow}= 0.944$ \\
\hline
\end{tabular}
\caption{Critical points, winding number and their correlation length critical exponents $\nu$ obtained by means of the penetration depth technique with $N=800$ for the edges of the rhombus (red circles) with $W=1$ at the topological phase diagram given by Fig.~\ref{fig:phase-diagram}. The $\nu_h$ (horizontal approach) and $\nu_v$ (vertical approach) as well as the arrows superscripted describe different forms to approach the transition point. We observe that all values of $\nu$ are very close to the unit.}
\label{tableI}\end{table}

On the other hand, for the transition within the diagonals of the rhombus with $W=1$, i.e., for the integer/semi-integer topological transitions, and vice versa, the ZEES exhibit distinct behaviors depending on the approach the transition point. For example, if we approach the transition point fixing the value of $v$ and increasing the degree of non-hermiticity of the system, the ZEES localize at the edges (left or right side) of the system due to the NH skin-effect~\cite{Yao-2018,Bergholtz-2021}, which is very similar to the behavior presented in Fig.~\ref{fig:semi-integer-trivial}~(b). By contrast, approaching the transition point fixing $v$ and decreasing the value of $g$ the ZEES tend to slightly penetrate into the bulk, which confirms that the degree of non-hermiticity of the system plays a important role to localize the ZEES at the edges (left or right side). Furthermore, if we approach the integer/semi-integer transitions along the left or right chiral region taking the same angular coeficient of the chiral lines in the topological phase diagram the ZEES are not affected, which denotes a non-critical feature for these type of NH topological transitions, due to the fact that there is no  diverging correlation length on this transition.

\section{Summary and discussions}
\label{Sec:discussions}

We have studied the NH topological phase transitions of the SSH model with asymmetric hopping using  numerical methods to compute the edge states and an analytical approach to display the topological characterization of the exceptional points. One of our main results is shown in Fig.~\ref{fig:phase_diagram_non_Herm_SSH}. We compute in detail the numerical ZEES and define the non-trivial chirality of them in the rhombus with winding $W = 1$. In particular, the line where $v = 0.0$ presents a Hermitian behavior due to the lack of the NH skin effect, i.e., each mode is on a specific edge of the system and, at the topological phase transitions, in $g = -1.0$ and $g=1.0$, the modes penetrate into the bulk. We use the flow of the exceptional points in the $\{\Re[h_x],\Re[h_y]\}$ space to describe the chiral behavior in the entire rhombus.

We also confirm that the degree of non-hermiticy of the NH systems is related to the skin-effect~\cite{Yao-2018,Bergholtz-2021} that localizes the edge states at the boundary (left of right side) of the system. Our numerical results show that the semi-integer/integer topological transition, and vice versa, presents an unusual aspect on the edge states, i.e., depending on the direction in parameter space in which the transition point is approached, the edge states may localize or might slightly penetrate into the bulk. Interestingly, when the transition point is approached considering the same angular coefficient of the chiral lines of the topological phase diagram,  these edge states are invariant at both sides of the transition. Thus, the transition from $W=1/2$ to $W=1$~(or vice versa), is not critical and the edge states remain equally localized at each side of the transition in the phase diagram.

From the complex gap structure analysis we confirm that for the Hermitian case, i.e., for $g=0.0$, the spectrum of the system does not change when the system undergoes a topological phase transition, exhibiting the real line gap, as expected. On the other hand, when we consider NH effects, that is, for $g \neq 0.0$, we show that one can identify all the NH topological phase transitions on the NH SSH model through the changing on its spectrum.

Moreover, applying the penetration depth technique~\cite{Griffith2018,Nei2019,Nei_2021},  we have obtained critical exponent ($\nu$) for the topological phase transitions at the vertex of  the rhombus with $W=1$ and in the transition lines $W= 1/2\to 0$, in the topological phase diagram of the NH SSH model. Our results show that the universality class of the NH SSH model is the same of the Hermitian SSH model for these specific transition points. 

It would be interesting to investigate the model in the limit $v \rightarrow 0$, where we have hermitian topological phase transitions even though the system is non-hermitian. Also the integer/semi-integer phase transition still represents a challenge. So, these aspects might be an interesting direction for further investigations.

\acknowledgments
The Brazilian agencies {\em Funda\c c\~ao Carlos Chagas Filho de Amparo \`a Pesquisa do Estado do Rio de Janeiro} (FAPERJ), {\em Conselho Nacional de Desenvolvimento Cient\'\i fico e Tecnol\'ogico} (CNPq) and {\em Coordena\c c\~ao  de Aperfei\c coamento de Pessoal de N\'\i vel Superior}  (CAPES) - Finance Code 001,  are acknowledged  for partial financial support. R.A. would like to thank FAPERJ for his current PhD fellowship. N.L. would like to thank the FAPERJ for the postdoctoral fellowship of the {\em Programa de Pós-Doutorado Nota 10 - 2020} (E-26/202.184/2020) as well as for the {\em Bolsa de Bancada para Projetos} (E-26/202.185/2020).


%

\end{document}